# Junction termination extension (JTE) with variation lateral doping (VLD) optimization method.

October 2011
E. Chernyavskiy, evgenyc@eclipso.ch

**Annotation.**

A simple and effective method for the junction termination design was suggested. Optimization method uses lateral charge function F(x) which depends from two arguments and can be changed in wide range of shapes. To demonstrate method effectiveness, design and optimization example for the HV diode (1800 V) edge termination was shown. Achieved breakdown voltage is 93% of the corresponding 1D structure breakdown voltage.

**Introduction.**

The breakdown voltage of a device is closely related to the maximum electric field at the junction of the devices. Increased electric field will significantly reduce the ability of power devices to withstand and block high reverse voltages. Electrical breakdown caused by impact ionization process will preferably occur at the junction periphery if the maximum electric field in these areas is not reduced by proper edge termination design. In fact, the edge termination limits the breakdown voltage to below the "ideal" limits set by corresponding 1D structure. For the real devices breakdown voltage can be as low as 10-20% of the ideal case.

Many designs have been employed to solve this periphery problem. Guard rings, field plates, argon implantation and Junction Termination Extensions (JTEs) have been used for planar power devices [1-4]. These methods have been successful for the most part, but each method has its particular drawbacks. Guard rings are often difficult design and optimize, field plates are limited by the strength of the dielectric used, argon implantation can increase reverse leakage current. Junction termination extensions have been widely used, but JTEs are difficult to optimize and often require multiple zones of decreasing implant dose in order to achieve ideal breakdown for a junction.

The JTE technique consists in extending the highly doped main junction by a connected surrounding region of the same type of conductivity but presenting a lower doping level, in order to allow the spreading of the equipotential lines emerging below the junction edge curvature towards the surface. Traditional JTE designs required precise control of dopant's in the JTE layer in order to completely deplete it at the desired blocking voltage. For a given doping and thickness of the substrate, and extension junction width are the main parameters affecting the blocking voltage.

To overcome JTE drawbacks a method of a varying lateral doping (VLD) was offered [5,7,8,9]. Screening mask for the JTE implantation designed as a series of window openings and spacings. That design allows to control amount of charge in lateral direction. A dopant is then implanted through the screening mask, after annealing forms the graded termination of the junction. The laterally varying charge of a JTE reduces field concentration at the p-n junction by spreading the field over a larger area, making the junction less vulnerable to high voltage breakdown and current leakage when reverse-biased.

JTE VLD design performs a significant problem in terms of optimization. In presented paper author suggest approach, which allows to significantly reduce design efforts.

**Ssingle zone JTE breakdown voltage simulation.**

As an example for JTE VLD design and optimization was used a punch-through (PT) diode. Silicon substrate thickness is 140 um, substrate doping $Nd=4E13$ cm$^{-3}$, N- buffer depth is 10 um, N- buffer surface doping density $Nds=3E15$ cm$^{-3}$. Diode net doping cross-section is shown at Figure 1. Simulated breakdown voltage for 1D structure is $Vbr=1993$ V, that corresponds a theoretical limit for a given semiconductor structure.

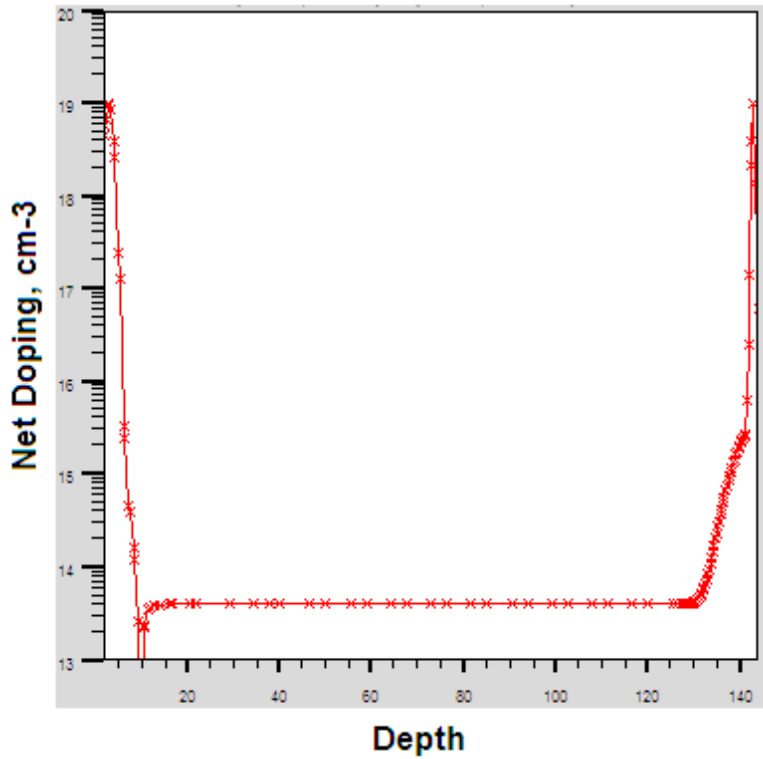

Figure 1. Diode net doping cross section.

As an initial approach the single zone JTE design was used for diode junction termination.

JTE layer parameters are: surface doping density range $Nas=3E15 - 5.5E15$ cm$^{-3}$, junction depth $Xj=7$ um, JTE width range $W=196 – 236$ um.

Total structure width, from diode P+ to the die edge is 400 um. Diode edge termination cross section with single zone JTE is shown at Figure 2.

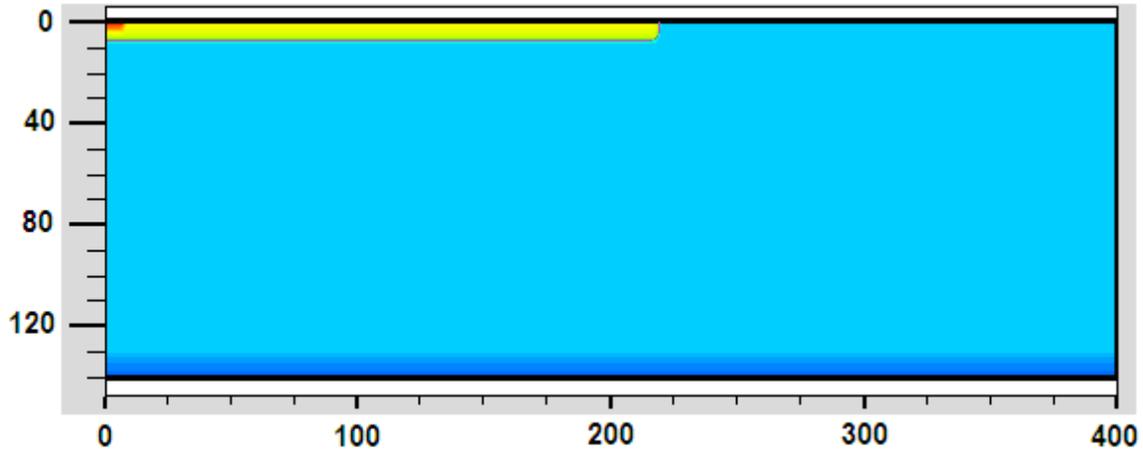

Figure 2. Diode edge termination cross section with single zone JTE, JTE width W=216 um.

Structure breakdown voltage with parameters variation is shown at Figure 3, all simulations were performed at temperature T=25 C.
Maximum breakdown voltage was found at JTE width W=236 um and surface boron density Nas=4E15 $cm^{-3}$. Simulation indicates that reduction of the breakdown voltage is more pronounced for with JTE doping variation towards higher doses.

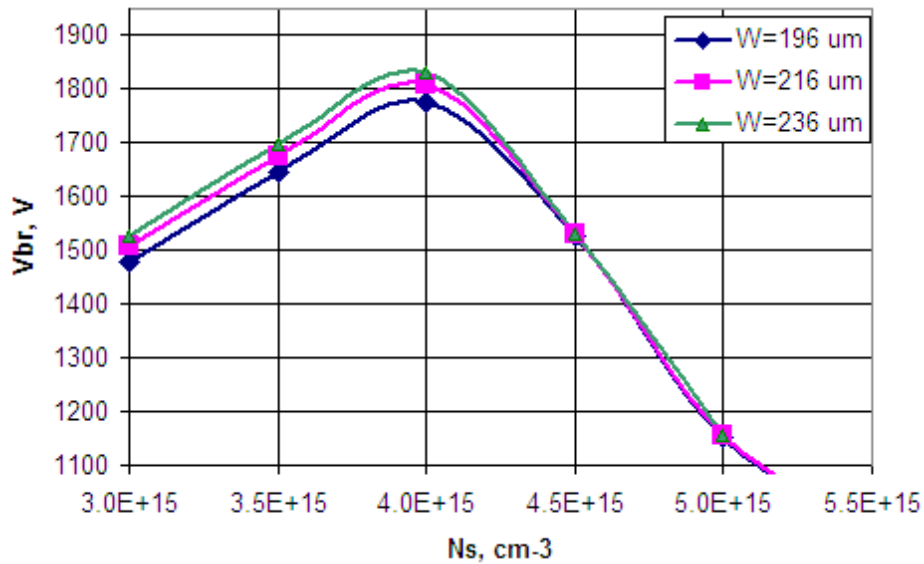

Figure 3. Single zone JTE breakdown voltage Vbr with respect to doping and width.

Although the ideal JTE charge will in theory give the maximum breakdown voltage, it has been shown in Si devices that including a fraction of the ideal charge, usually 60-80% of Qi [10], can reduce high surface fields. These high surface fields can lead to a number of detrimental effects in power devices, including: excess leakage, breakdown walkout, and premature breakdown at the surface of the device due to trapped charge or defects. Therefore, including only 80% of the optimum JTE dose should give a good trade-off between the blocking capabilities and the reduction of surface field. To improve that trade-off Variation Lateral Doping (VLD) structure was proposed. VLD structure allow to control of surface peak field without compromising the breakdown voltage.

**Lateral charge profile function.**

The problem of finding optimal lateral charge profile relates to the type of reverse problems of mathematical physics. Method of solution, used in current paper is in follows: first, a breakdown voltage simulated for a given lateral charge profile. Then with some alterations in doping profile a breakdown voltage simulated again, ant that repeats several times. That generates a grid of possible solutions. Using that set of solutions optimal lateral charge profile can be found. To procedure simplify a lateral function with two arguments B and C was suggested (Formula 1). Arguments B and C variations can cover a wide range of possible lateral doping distributions.

$$F(x)=0.5\cdot\left[\frac{(1-\exp(B\cdot(x-C)))}{(1+\exp(B\cdot(x-C)))}+1\right] \quad (1)$$

To design a screening mask with window openings and spacing, lateral function F(x) has to be performed as a discrete function implantation window openings and spacings. Algorithm for such discretization are similar that used for Pulse-Width Modulation (PWM). Function F(x) with arguments B=0.03, C=140 and corresponding mask openings and spacings is shown at Figure 4.

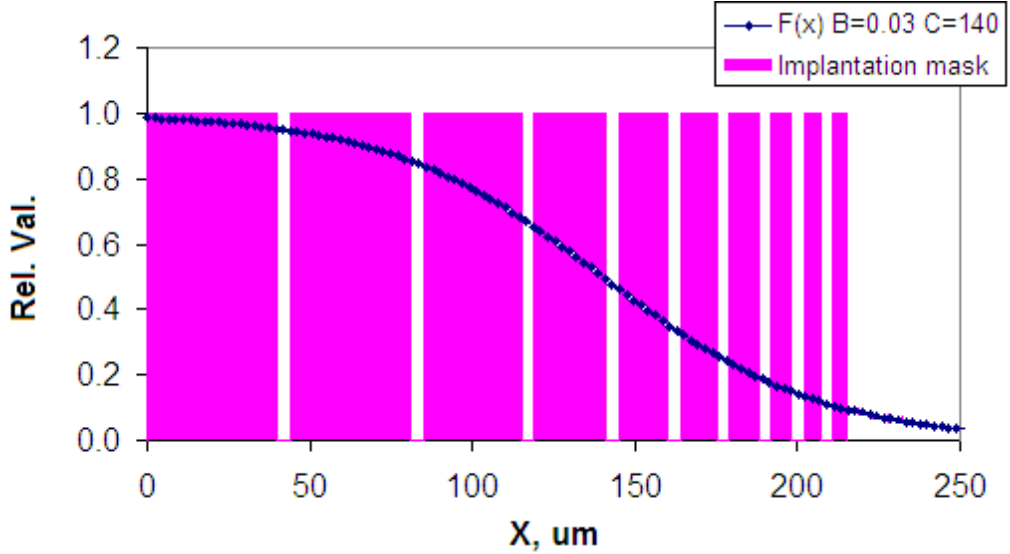

Figure. 4. Lateral charge function F(x) (B=0.03, C=140) and its discretization as mask for the implantation.

**JTE VLD breakdown voltage simulation.**

Since a reduction of JTE charge to decrease the peak surface electric field would result in a significant blocking capability loss, the Variation Lateral Doping (VLD) was proposed to allow control of surface peak field without compromising the breakdown voltage.

JTE VLD structure breakdown voltage was simulated with variation in lateral charge, controlled by lateral charge distribution function F(x) described above. Function F(x) arguments B and C was varied and corresponding implantation mask design was used for breakdown voltage simulation, with surface boron density Nas= 4E15 cm$^{-3}$. Simulation results are shown at the Table 1.

**Table 1. Simulated breakdown voltages as a function of arguments B and C.**

|       | B=0.003   | B=0.01    | B=0.03    | B=0.06    |
|-------|-----------|-----------|-----------|-----------|
| C=140 | Vbr=1835 V | Vbr=1861 V | Vbr=1850 V | Vbr=1833 V |
| C=160 | Vbr=1838 V | Vbr=1838 V | Vbr=1844 V | Vbr=1847 V |
| C=180 | Vbr=1840 V | Vbr=1848 V | Vbr=1862 V | Vbr=1826 V |
| C=200 | Vbr=1840 V | Vbr=1854 V | Vbr=1871 V | Vbr=1863 V |

That simulations indicates some improvements in breakdown voltages, but improvements in reusing surface field is still unclear. To compare width of the dose tolerance window between single zone JTE and JTE VLD structures, additional simulations were performed. Surface doping density was varied for lateral charge function F(x) with arguments B=0.03, C=200 and B=0.03, C=140, and breakdown voltage simulated (Figure 5).

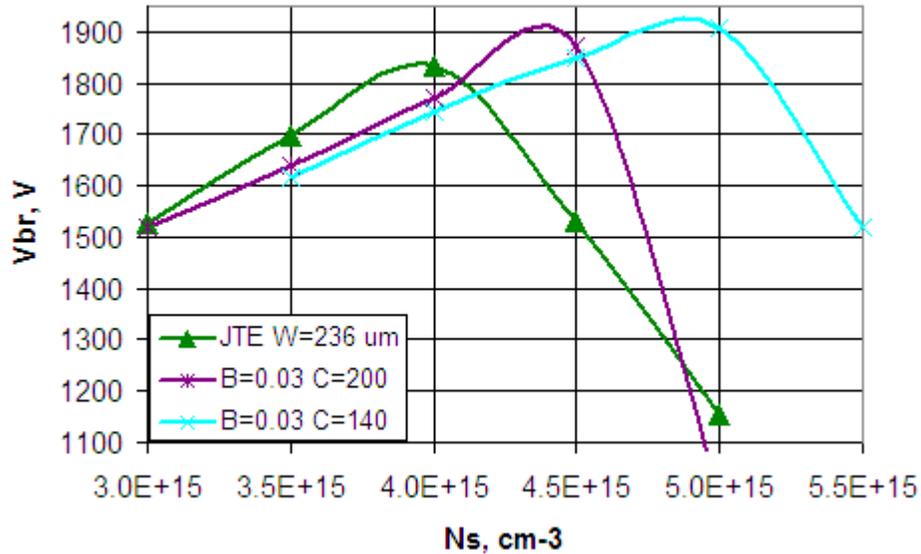

**Figure 5. Dose tolerance window comparison for the single zone JTE and JTE VLD with different lateral charge distribution.**

As it can be seen from the Figure 5, in case of JTE VLD dose tolerance window is in factor of 2.8 wider than for the single zone JTE. That provides more possibilities in terms of reducing surface field by using less charge regarding ideal once.

**Conclusions.**

A simple and effective method for the junction termination design was suggested. Optimal solution for the Junction Termination Extension with Variation Lateral Doping (JTE VLD) can be found within several days of simulations. Optimization method uses lateral charge function F(x) which depends from two arguments and can be changed in wide range of shapes.
To demonstrate method effectiveness, design and optimization example for the HV diode (1800 V) edge termination was shown. Achieved breakdown voltage is 93% of the corresponding 1D structure breakdown voltage.

**References.**